\documentstyle[aps]{revtex}
\def\vec#1{\mathchoice
 {\mbox{\boldmath $\displaystyle#1$}}
 {\mbox{\boldmath $\textstyle#1$}}
 {\mbox{\boldmath $\scriptstyle#1$}}
 {\mbox{\boldmath $\scriptstyle#1$}}}
\begin{document}

\title{Nuclear polarization in hydrogenlike 
$^{208}_{~82}$Pb$^{81+}$}
\author{Akihiro Haga$^{1}$
\footnote{Electronic address: haga@npl4.kyy.nitech.ac.jp}
, Yataro Horikawa$^2$, 
\footnote{Electronic address: horikawa@sakura.juntendo.ac.jp}
and 
Yasutoshi Tanaka$^1$ \\[1ex]
\it
$^1$\,Department of  Environmental Technology and Urban Planning, \\
\it
Nagoya Institute of Technology,
Gokiso, Nagoya 466-8555, Japan\\[1ex]
$^2$\,Department of Physics, Juntendo University, Inba-gun, Chiba
 270-1695, Japan
\it
}
\date{5 October 2001}
\maketitle
\setlength{\baselineskip}{0.3in}
\vspace{5mm}

\begin{abstract}
\setlength{\baselineskip}{0.3in}
We calculate nuclear-polarization energy shifts
for hydrogenlike $^{208}_{~82}$Pb$^{81+}$. A retarded 
transverse part as well as the Coulomb part is taken 
into account as the electromagnetic interaction 
between an electron and the nucleus. 
With a finite charge distribution for the nuclear 
ground state and the random-phase approximation 
to describe the nuclear excitations, we obtain 
nuclear polarization energy of the $1s_{1/2}$ state 
as --38.2 (--37.0) meV in the Feynman (Coulomb) gauge. 
For the $2s_{1/2}$, $2p_{1/2}$ and $2p_{3/2}$ states, 
they are --6.7 (--6.4), --0.2 (--0.2) and +0.0 (+0.0) meV, 
respectively. The seagull term in the two-photon
exchange diagrams is shown to be quite important to obtain
the gauge invariance of nuclear polarization energies.
\\
\\
PACS number(s): 31.30.Gs, 31.30.Jv, 12.20.Ds
\end{abstract}

\section{Introduction}

High-precision Lambshift measurement on high-Z 
hydrogenlike atoms\cite{BE95} arose a renewed 
interest in the quantum electrodynamic
(QED) calculation of electronic atoms. 
Comparison of theoretical results with corresponding 
experimental data allows sensitive tests of
QED in strong electromagnetic fields \cite{SA90,MO98}. 
Any discrepancy between theory and experiment 
may either motivate an improvement of theoretical 
calculations and a refinement of experiments, or it may
indicate a possible influence of non-QED effects. 
In this context, the study of nuclear polarization (NP) 
contributions to the total energy shift of atomic levels 
becomes important because as a background effect, 
it represents a natural limitation of any
high-precision test of QED. Unfortunately, 
evaluation of NP is not practicable on the first principle. 
Any calculation of NP is inherently phenomenological 
and depends on the parameters of the nuclear model 
used to describe the intrinsic nuclear dynamics.

During the past years, a lot of experience has been 
accumulated in calculating the NP effect for muonic 
atoms\cite{RI82}. 
There it leads to a large correction at a keV level, 
mainly because of the huge overlap of the
muon-wave function with a nucleus and because the transition
energies in muonic atoms are of the order of 
magnitude of typical nuclear excitation energies.

Much less attention has been paid to the NP effect 
for electronic atoms. 
They turn out to be reduced by orders of magnitude 
because of the small overlap of the electron-wave 
function with the nucleus and because the transition 
energies in electronic atoms are in general orders of 
magnitude smaller than typical nuclear excitation energies.

The NP effect for electronic atoms was first calculated 
in terms of the second-order Schr\"{o}dinger perturbation 
theory \cite{HO84}. 
A relativistic field-theoretical treatment of 
nuclear polarization calculation was then presented 
by Plunien et al. \cite{PL89,PL95,NE96} 
utilizing the concept of effective photon propagators 
with nuclear polarization insertions. 
The formalism allows to take into account the
effect of the electron negative-energy intermediate 
states besides the usual contribution of the electron 
excited into higher unoccupied intermediate states. 
They found that in electronic atoms NP energies
become small due to the cancellation between contributions of 
positive energy states and those of negative energy states.

In the above studies, only the Coulomb interaction
was considered based on the argument that the relative 
magnitude of transverse interaction is of the order 
of $ (v/c)^2$ and the velocity $v$ associated with 
nuclear dynamics is mainly nonrelativistic. 
However, it may not be justified. In fact the
importance of the transverse interaction has been 
reported for muonic atoms \cite{CO69,TA94}. 
In Ref.~\cite{TA94}, the transverse nuclear
polarization has been studied in order to explain 
the discrepancies between theory and experiment 
in the 2p and 3p fine-structure splitting energies 
of muonic $^{208}_{~82}$Pb. 
The contribution for the muonic $1s_{1/2}$ state 
amounts to 20\% of that of the Coulomb interaction.

The transverse interaction is expected to be much 
more important for electronic atoms than for muonic 
atoms because of its long-range nature.  
The transverse nuclear polarization for heavy electronic
atoms was first studied by Yamanaka et al.\cite{YA99,YA01} 
using the Feynman gauge and a collective model 
for the nuclear excitations. 
They found that the transverse contribution is several
times larger than the Coulomb contribution in heavy 
electronic atoms before the contributions of the 
positive and negative energy states cancel. 
However, due to nearly complete cancellation between
them, the total NP energy becomes very small.

The purpose of the present paper is twofold; 
The one is to see how much NP energies are expected 
for a practically best available model
of the $^{208}_{~82}$Pb nucleus. 
For this purpose the Dirac-electron wave functions are 
solved in the Coulomb potential with a finite nuclear 
charge distribution and the random-phase approximation
(RPA) is used to describe the nuclear excitations. 
The other is to see whether NP energies are sensitive 
to the choice of the gauge. 
For this purpose NP energies are calculated both in the 
Feynman and Coulomb gauges.
We will see that NP calculation with only the ladder 
and cross diagrams shows large gauge dependence and 
inclusion of the seagull diagram removes most of its 
gauge dependence\cite{FR74,RO83}.

Calculations are carried out in momentum space.
They involve only double integrals,
which are easily carried out with high precision.

\section{nuclear polarization calculation}

The second-order contributions to the nuclear polarization  
are given by three Feynman diagrams in Fig.~1. 
(Here we regard the seagull graph as one of the
nuclear polarization diagrams.) 
Two photons are exchanged between a bound electron and a nucleus 
and the nuclear vertices are understood to have no 
diagonal matrix elements for the ladder and cross diagrams,
and no nuclear intermediate states for the seagull diagram.

The nuclear-polarization energy shift due to the ladder 
and cross diagrams is given by \cite{PL89},
\begin{eqnarray}
\Delta E_{NP}= i (4\pi\alpha)^2
\int d^4x_1\cdots d^4x_4 &&
\bar{\psi}(x_1)\gamma^{\mu}S^e_F(x_1,x_2)\gamma^{\nu}\psi(x_2)
\nonumber
\\
&&\times 
D_{\mu\xi}(x_1,x_3)
\Pi^{\xi\zeta}_N(x_3,x_4)\,
D_{\zeta\nu}(x_4,x_2).
\label{DENP1}
\end{eqnarray}
Here $\psi$ is the electron wave function, $S^e_F$ 
the external-field electron propagator, 
$D_{\mu\xi}$ the photon propagator and $\Pi^{\xi\zeta}_N$ 
is the nuclear polarization tensor which contains
all information of nuclear dynamics. We use units with
$\hbar=c=1$ and $e^2=4\pi\alpha$.

In terms of transition charge-current densities, 
the electron and nuclear parts of Eq.~(\ref{DENP1}) 
are written as
\begin{equation}
\bar{\psi}(x_1)\gamma^{\mu}S^e_F(x_1,x_2)\gamma^{\nu}
\psi(x_2)=\int\frac{dE}{2\pi}e^{-i E(t_1-t_2)}\sum_{i'}\frac
{j^{\mu}_e(\vec{x}_1)_{ii'}j^{\nu}_e(\vec{x}_2)_{i'i}}
{E-\omega_e+i E_{i'}\epsilon}
\label{electron}
\end{equation}
and
\begin{equation}
\Pi^{\xi\zeta}_N(x_3,x_4)=
\int\frac{d\omega}{2\pi}e^{-i \omega 
(t_3-t_4)}\sum_{I'}\left(\frac{J^{\xi}_{N}(\vec{x}_3)_{II'}
J^{\zeta}_{N}(\vec{x}_4)_{I'I}}
{\omega-\omega_N+i \epsilon}
-\frac{J^{\zeta}_{N}(\vec{x}_4)_{II'}J^{\xi}_{N}
(\vec{x}_3)_{I'I}}{\omega+\omega_N-i \epsilon}\right),
\label{nuclus}
\end{equation}
where  $\omega_e=E_{i'}-E_i$  and $\omega_N=E_{I'}-E_I$ are 
excitation energies of electron and
nucleus, respectively.
The suffixes $i(I)$ and $i'(I')$ 
stand for the initial and intermediate
states of the electron (nucleus), respectively.

In the momentum representation, the NP  energy shifts 
due to the ladder and cross diagrams are written as
\begin{eqnarray}
\Delta E_{NP}^L&=&
-i (4\pi\alpha)^2\int \frac{d\omega}{2\pi}\int
\frac{d\vec{q}}{(2\pi)^3}
\int \frac{d\vec{q}'}{(2\pi)^3}
D_{\mu\xi}(\omega,\vec{q})D_{\zeta\nu}(\omega,\vec{q}')
\nonumber
\\
\nonumber
\\
&&\times
\sum_{i'}\frac{j_e^{\mu}(-\vec{q})_{ii'}j_e^{\nu}
(\vec{q}')_{i'i}}
{\omega+\omega_e-i E_{i'}\epsilon}
\sum_{I'}
\frac{J_N^{\xi}(\vec{q})_{II'}J_N^{\zeta}(-\vec{q}')_{I'I}}
{\omega-\omega_N+i \epsilon}
\label{DENPL}
\end{eqnarray}
and
\begin{eqnarray}
\Delta E_{NP}^X&=&
i (4\pi\alpha)^2\int \frac{d\omega}{2\pi}\int
\frac{d\vec{q}}{(2\pi)^3}
\int \frac{d\vec{q}'}{(2\pi)^3}
D_{\mu\xi}(\omega,\vec{q})D_{\zeta\nu}(\omega,\vec{q}')
\nonumber
\\
\nonumber
\\
&&\times
\sum_{i'}\frac{j_e^{\mu}(-\vec{q})_{ii'}j_e^{\nu}
(\vec{q}')_{i'i}}
{\omega+\omega_e-i E_{i'}\epsilon}
\sum_{I'}
\frac{J_N^{\zeta}(-\vec{q}')_{II'}J_N^{\xi}(\vec{q})_{I'I}}
{\omega+\omega_N-i \epsilon}, 
\label{DENPX}
\end{eqnarray}
respectively. 
The substitution
\begin{equation}
\Pi^{\xi\zeta}(x_3,x_4)\rightarrow 
\frac{\rho_{N}({\vec x}_3)_{II}}{m_p}
\delta^{\xi\zeta}\delta^4(x_3-x_4)
\label{SEA}
\end{equation}
in Eq.~(\ref{DENP1}) gives the energy correction 
due to the seagull diagram
\cite{FR74,RO83}
\begin{eqnarray}
\Delta E_{NP}^{SG}&=&
-i (4\pi\alpha)^2\int \frac{d\omega}{2\pi}\int
\frac{d\vec{q}}{(2\pi)^3}
\int \frac{d\vec{q}'}{(2\pi)^3}
D_{\mu\xi}(\omega,\vec{q})D_{\zeta\nu}(\omega,\vec{q}')
\nonumber
\\
\nonumber
\\
&&\times
\sum_{i'}\frac{j_e^{\mu}(-\vec{q})_{ii'}j_e^{\nu}
(\vec{q}')_{i'i}}
{\omega+\omega_e-i E_{i'}\epsilon}
\frac{\rho_{N}(\vec{q}-\vec{q}')_{II}}{m_p}\delta^{\xi\zeta}.
\label{DENPSG}
\end{eqnarray}
Here $m_p$ is the proton mass, $\delta^{\xi\zeta}$ 
the Kronecker delta extended to four dimensions 
with $\delta^{00}=0$, and $\rho_{N}(\vec{x})_{II}$ 
is the ground-state charge distribution of the nucleus. 
The total NP energy shift is given by the sum, 
$\Delta E_{NP}^L + \Delta E_{NP}^X + \Delta E_{NP}^{SG}$.

Substituting the corresponding multipole 
expansions for the Fourier
transforms of the currents,  
these NP energy shifts 
are  written in terms of the multipole form factors
of the electron and nucleus defined by
\begin{eqnarray}
<i'\parallel m_{\lambda}(q)\parallel i>&=&\int d{\vec x}
\; j_\lambda(qx)
<i'\parallel Y_\lambda(\Omega_x)\rho_e({\vec x})\parallel i>,
\label{echarge}
\\
<i'\parallel t_{\lambda L}(q)\parallel i>&=&\int d{\vec x}
\;j_L(qx)
<i'\parallel {\vec Y}_{\lambda L}(\Omega_x)
\cdot{\vec j}_e({\vec x})\parallel i>
\label{ecurrent},
\end{eqnarray}
for the electron, and 
\begin{eqnarray}
<I'\parallel M_{\lambda}(q)\parallel I>&=&\int d{\vec x}
\;j_\lambda(qx)
<I'\parallel Y_\lambda(\Omega_x)\rho_N({\vec x})\parallel I>,
\label{ncharge}
\\
<I'\parallel T_{\lambda L}(q)\parallel I>&=&\int d{\vec x}
\;j_L(qx)
<I'\parallel {\vec Y}_{\lambda L}(\Omega_x)
\cdot{\vec J}_N({\vec x})\parallel I>
\label{ncurrent},
\end{eqnarray}
for the nucleus. In the above equations, $j_\lambda(qx)$ 
is a spherical Bessel function, 
$\vec{Y}_{\lambda L}$ is a vector spherical
harmonics and $\lambda$ is the multipolarity of transition. 
With these substitutions, angular parts of $\vec{q}$ 
and $\vec{q}'$ as well as $\omega$ integrations 
in Eqs.~(\ref{DENPL}), (\ref{DENPX}) and
(\ref{DENPSG}) can be carried out analytically. 
Integrations with respect to $q$ and $q'$ are 
carried out numerically. 
Nuclear-polarization energy shifts are thus given by 
the sum of these double integrals over the nuclear and 
electron intermediate states. 
For the seagull contribution, the summation is only 
over the electron intermediate states.

In the following, we shall give formula for the NP energy 
both in the Feynman and Coulomb gauges. 
We restrict ourselves to $I^\pi=0^+$ for
the spin-parity of the nuclear ground state. 
The spin-parity of the nuclear intermediate state 
$I'^\pi$ is in this case equal to the
spin-parity of the transition $\lambda^\pi$.

\subsection{The Feynman gauge}

The photon propagator in the Feynman gauge is given by
\begin{equation}
D_{\mu\xi}^F(\omega,\vec{q})=-\frac{g_{\mu\xi}}
{q^2+i \epsilon}.
\label{DFeynman}
\end{equation}
Here the metric tensor $g_{\mu\xi}$ is defined by $g_{00}=1$ 
and $g_{ii}=-1$. With this propagator, the NP energies 
due to the ladder,
cross and seagull terms are given by
\begin{eqnarray}
\Delta E_{NP}^L&=& - \sum_{i'I'}
\frac{(4\pi\alpha)^2}{(2i+1)(2 I'+ 1)}
(\frac{2}{\pi})^2
{\cal P}
\int^{\infty}_{0}\!\!dq \int^{\infty}_{0}\!\!dq'
\ I_+ (q,q')\ {\cal W}^L_F(q){\cal W}^L_F(q')\delta_{I'\lambda},
\label{DENPFEL}\\
\nonumber
\\
\Delta E_{NP}^X&=& - \sum_{i'I'}
\frac{(4\pi\alpha)^2}{(2i+1)(2 I' + 1)}
(\frac{2}{\pi})^2
{\cal P}
\int^{\infty}_{0}\!\!dq \int^{\infty}_{0}\!\!dq'
\ I_-(q,q')\ {\cal W}^X_F(q){\cal W}^X_F(q')\delta_{I'\lambda},
\label{DENPFEX}
\\
\nonumber
\\
\Delta E_{NP}^{SG}&=& -\sum_{i'}\frac{(4\pi\alpha)^2}{(2i+1)}
(\frac{2}{\pi})^2{\cal P}\int^{\infty}_{0}\!\!dq 
\int^{\infty}_{0}\!\!dq'
\ I_{SG}(q,q')\ {\cal W}^F_{SG}(q,q'),
\label{DENPFESG}
\end{eqnarray} 
where ${\cal P}$ denotes
the Cauchy principal value of any improper
integral. The $q$ and $q'$ integrals may be improper 
in case of the NP calculation for the electron excited states.

In the above expressions, $I_+, I_-$ and $I_{SG}$ are 
functions written as  
\begin{eqnarray}
I_{\pm}(q,q')&=& {qq' \over
2 (\tilde{\omega}_e+q)(\tilde{\omega}_e+q')
(\omega_N+q)(\omega_N+q')}
\\
\nonumber
&&\times
\left\{{\rm sgn}(E_{i'})[{\tilde{\omega}_e\omega_N \over q+q'}
+(\tilde{\omega}_e+\omega_N+q+q')] 
\pm {qq' \over q + q'} \pm 
\theta(\pm E_{i'}) {2qq' \over \omega_N+ 
\tilde{\omega}_e}\right\},
\label{eq:D_ell}
\\
I_{SG}(q,q')&=& {\rm sgn}(E_{i'}) {1 \over 2m_p}
\frac{qq'(\tilde{\omega}_e+q+q')}
{(q+q')(q+\tilde{\omega}_e)
(q'+\tilde{\omega}_e)},
\label{eq:D_ellM}
\end{eqnarray}
where $\tilde{\omega}_e={\rm sgn}(E_{i'})\omega_e$,
while ${\cal W}^F_L(q)$, 
${\cal W}^F_X(q)$ and ${\cal W}^F_{SG}(q)$ are
written in terms of the electron and nuclear form factors as
\begin{eqnarray}
{\cal W}^F_L(q)&=& \sum_\lambda 
[ <i' \parallel m_{\lambda}(q)\parallel i>
<I' \parallel M_{\lambda}(q)\parallel I> 
\nonumber
\\
&-& \sum_{L=\lambda-1}^{\lambda+1}(-1)^{L+1-\lambda}
<i' \parallel t_{\lambda L}(q)\parallel i>
<I' \parallel T_{\lambda L}(q)\parallel I>],
\\
{\cal W}^F_X(q)&=& \sum_\lambda [ 
<i' \parallel m_{\lambda}(q)\parallel i>
<I' \parallel M_{\lambda}(q)\parallel I> 
\nonumber
\\
&+& \sum_{L=\lambda-1}^{\lambda+1}
<i' \parallel t_{\lambda L}(q)\parallel i>
<I' \parallel T_{\lambda L}(q)\parallel I>],
\\
{\cal W}^F_{SG}(q,q')&=&\sum_\lambda 
\sum_{L=\lambda-1}^{\lambda+1}
<i' \parallel t_{\lambda L}(q)\parallel i>
<I \parallel M_{L}(q,q')\parallel I>
<i' \parallel t_{\lambda L}(q')\parallel i>.
\label{FEMSG}
\end{eqnarray}
The seagull term contains the Fourier transform of 
the nuclear ground state,
\begin{eqnarray}
<I \parallel M_{L}(q,q')\parallel I>=
\int r^2 dr \rho_N(r)_{II}j_{L}(qr)j_{L}(q'r).
\label{ML}
\end{eqnarray}

\subsection{The Coulomb gauge}

The photon propagator in the Coulomb gauge is 
given by
\begin{equation}
D^C_{00}(\omega,\vec{q})=\frac{1}{q^2+i\epsilon},
\hskip1cm
D^C_{ij}(\omega,\vec{q})=\frac{1}{q^2+i\epsilon}
(\delta_{ij}-\frac{q_iq_j}{|\vec{q}|^2})
\label{DCoulomb},
\end{equation}
and $(\delta_{ij}-\frac{q_iq_j}{|\vec{q}|^2})$ 
in $D^C_{ij}$ projects out
both transverse parts of electronic and nuclear currents.

Making use of the relation 
\begin{equation}
\vec{j}_e^T(-\vec{q})\cdot\vec{J}_N^T(\vec{q})=
\vec{j}_e(-\vec{q})\cdot
\vec{J}_N(\vec{q})+\frac{\omega_e \omega_N}{q^2}
\rho_e(-\vec{q})
\rho_N(\vec{q}),
\label{PROTCUR}
\end{equation}
we obtain the NP energies for the ladder, cross 
and seagull terms:
\begin{eqnarray}
\Delta E_{NP}^L&=& - \sum_{i'I'}\frac{(4\pi\alpha)^2}
{(2i+1)(2I'+1)}
{\cal P}\int^{\infty}_{0}\!\!dq \int^{\infty}_{0}\!\!dq'
\left\{{\theta(E_{i'}) \over \tilde{\omega}_e+\omega_N}
{\cal W}^C_L(q){\cal W}^C_L(q') \right.
\nonumber
\\
\nonumber
\\
&+&\left. I_+(q,q'){\cal W}^C_T(q){\cal W}^C_T(q')+I_+^{LT}(q') 
{\cal
W}^C_L(q){\cal W}^C_T(q')\frac{}{}\right\}\delta_{I'\lambda},
\label{DENPCOL}
\\
\nonumber
\\
\Delta E_{NP}^X&=& - \sum_{i'I'}\frac{(4\pi\alpha)^2}
{(2i+1)(2I'+1)}
{\cal P}\int^{\infty}_{0}\!\!dq \int^{\infty}_{0}\!\!dq'
\left\{\frac{- \theta(-E_{i'})}{\tilde{\omega}_e+\omega_N}
{\cal W}^C_L(q){\cal W}^C_L(q') \right.
\nonumber
\\
\nonumber
\\
&+&\left. I_-(q,q'){\cal W}^C_T(q){\cal W}^C_T(q')
+I_-^{LT}(q') {\cal W}^C_L(q){\cal W}^C_T(q')
\frac{}{}\right\}\delta_{I'\lambda}, 
\label{DENPCOX}
\\
\nonumber
\\
\Delta E_{NP}^{SG}&=&- \sum_{i'}
\frac{(4\pi\alpha)^2}{(2i+1)}(\frac{2}{\pi})^2
{\cal P}\int^{\infty}_{0}\!\!dq \int^{\infty}_{0}\!\!dq'
I_{SG}(q,q'){\cal W}^C_{SG}(q,q').
\label{DENPCOSG}
\end{eqnarray}

In the above expressions, $I_+^{LT}(q')$ and $I_-^{LT}(q')$  
come from the interference between the longitudinal and 
transverse contributions and given by
\begin{equation}
I_\pm^{LT}(q')  = \pm\frac{{\mathrm sgn}(E_{i'}) 
q'(\tilde{\omega}_e+\omega_N)
\pm \theta(\pm E_{i'}) 2q'^2}
{(q'+\tilde{\omega}_e)(q'+\omega_N)
(\tilde{\omega}_e+\omega_N)}.
\end{equation}

In the Coulomb gauge, ${\cal W}^C_L(q)$, ${\cal W}^C_L(q)$ 
and ${\cal W}^C_L(q)$ are given by
\begin{eqnarray}
{\cal W}^C_L(q)&=& \sum_\lambda 
<i' \parallel m_{\lambda}(q)\parallel i>
<I' \parallel M_{\lambda}(q)\parallel I>,
\\
\nonumber
\\
{\cal W}^C_T(q)&=& - \sum_\lambda 
\left[\frac{\omega_e \omega_N}
{q^2}{\cal W}^C_L(q) + \sum_{L=\lambda-1}^{\lambda+1}
<i' \parallel t_{\lambda L}(q)\parallel i>
<I' \parallel T_{\lambda L}(q)\parallel I>\right],
\\
\nonumber
\\
{\cal W}^C_{SG}(q,q')&=& \sum_\lambda 
\left[ \sum_{L = \lambda \pm 1} (<i' \parallel u_{\lambda L}
(q)\parallel i>
<I \parallel M_{L}(q,q')\parallel I>
<i' \parallel u_{\lambda L}(q')\parallel i>)\right. 
\nonumber
\\
\nonumber
\\
&+&\left. <i' \parallel t_{\lambda \lambda}(q)\parallel i>
<I \parallel M_{\lambda}(q,q')\parallel I>
<i' \parallel t_{\lambda \lambda}(q')\parallel i> 
\frac{}{}\right],
\label{sgc}
\end{eqnarray}
with
\begin{eqnarray}
  <i' \parallel u_{\lambda \lambda -1 }(q)\parallel i> &=&
<i' \parallel t_{\lambda-1}(q)\parallel i>
-i \sqrt{\frac{\lambda}{2\lambda+1}}
\frac{\omega_e}{q}
<i' \parallel m_{\lambda}(q)\parallel i>,
\\
  <i' \parallel u_{\lambda \lambda + 1 }(q)\parallel i> &=&
<i' \parallel t_{\lambda+1}(q)\parallel i>
-i \sqrt{\frac{\lambda+1}{2\lambda+1}}
\frac{\omega_e}{q}
<i' \parallel m_{\lambda}(q)\parallel i>.
\end{eqnarray}

\subsection{Electron wave functions}

 The radial Dirac equations for the electron are written as 
\begin{eqnarray}
\left(\frac{d}{dr}+\frac{\kappa}{r}\right)G_{E,\kappa}
&=&[m_e+E-V(r)]F_{E,\kappa}(r),
\\
\nonumber
\\
\left(\frac{d}{dr}-\frac{\kappa}{r}\right)F_{E,\kappa}
&=&[m_e-E+V(r)]G_{E,\kappa}(r),
\end{eqnarray}
where the potential $V(r)$ is obtained from the 
the ground-state charge 
distribution of $^{208}$Pb, which is assumed to be a two 
parameter Fermi distribution
\begin{equation}
\rho_N(r)_{II}=\frac{\rho_0}{1+\exp[(r-R_0)/a]}
\label{FermiD}
\end{equation}
with $R_0=6.6477$ fm and $a=0.5234$ fm\cite{TA94}.
These equations are solved numerically by using the
fourth-order Runge-Kutta method.
For both positive and negative energy continuum states, 
the radial functions are normalized
by the conditions
\begin{eqnarray}
G_{E,\kappa}&\mathop{\longrightarrow}
\limits_{r\rightarrow\infty}&
\left(\frac{|E+m_e|}{\pi p}\right)^{1/2}\sin(pr+\delta),
\\
\nonumber
\\
F_{E,\kappa}&\mathop{\longrightarrow}
\limits_{r\rightarrow\infty}&
\left(\frac{|E-m_e|}{\pi p}\right)^{1/2}\cos(pr+\delta),
\end{eqnarray}
with $p=\sqrt{E^2-m_e^2}$.  
 The bound-state wave functions are normalized as 
$[\int^\infty_0(G_{n\kappa}^2+F_{n\kappa}^2)dr=1]$.

Transition form factors for the electron, (\ref{echarge}) and
(\ref{ecurrent}), are calculated using the formula given in
Ref.~\cite{TA94}. 
They are stored in the computer with six different
step sizes of $\Delta q$ depending on the electron energy 
$E_{i'}$. In Fig.~2 we show the $E1$ charge form factors
$<E_{i'},p_{1/2}\parallel m_1(q)\parallel 1s_{1/2}>$ 
of the electron with three different energies, 
$E_{i'}$=2, 6 and 10 MeV. One finds that
they have sharp peaks at $q = E_{i'}$ and decrease 
rapidly as $q$ increases. 
Numerical integrations 
in Eqs.~(\ref{DENPFEL})--(\ref{DENPFESG}) and
(\ref{DENPCOL})--(\ref{DENPCOSG}) are performed 
by Simpson's one-third rule.

Most of the NP correction comes from the continuum 
states with energies greater than $m_e$ and energies 
less than $-m_e$. Summation
over the electron states $i'$ in
Eqs.~(\ref{DENPFEL})--(\ref{DENPFESG}) and
Eqs.~(\ref{DENPCOL})--(\ref{DENPCOSG}) implies an 
integration with respect to $E_{i'}$. 
The $E_{i'}$ integration is carried out by using the 
Gauss-Legendre quadrature over the intervals 
--250MeV$<E_{i'}<-m_e$ and $m_e<E_{i'}<$250MeV.
Accuracy of the numerical results is checked by
comparing them with the results of Simpson's rule. 
We also included the electron bound states 
for n'$\leq$7 in the calculation.

\subsection{RPA calculation of nuclear charge 
and current densities}

The random-phase approximation is used to describe the nuclear
excitations. The calculation we employed is the same as 
those performed earlier in Ref.~\cite{TA94}, i.e., 
the same single-particle basis, the
same particle-hole configuration of approximately a full
3$\hbar\omega$ space and the same Migdal force\cite{MI67}
parameters to describe
nuclear two-body  interaction. Nuclear transition form factors 
, (\ref{ncharge}) and (\ref{ncurrent}),
are calculated by using the formula given in Ref.~\cite{TA94}.

The calculated charge and magnetic current densities are 
examined by comparing with experimental $B(E\lambda)$ and
$B(M\lambda)$ using the following relations; 
\begin{equation}
B(E\lambda:I \rightarrow I')={1 \over 2I+1}\left\vert\,e\int
\rho^{I'I}_{N\lambda}(r)r^{\lambda+2}dr\right\vert^2
\end{equation}
and
\begin{equation}
B(M\lambda:I\rightarrow I')={1 \over 2I+1}
{\lambda\over{\lambda+1}}
\left\vert\,e\int J^{I'I}_{N\lambda\lambda}
(r)r^{\lambda+2}dr\right
\vert^2.
\end{equation}

In Table I, we compared the energy-weighted sum of
$B(E\lambda)$ over the RPA states with the classical 
energy-weighted sum-rule value (EWSR) of Ref.~\cite{BM75}. 
The results for the $E0$ and $E1$
transitions exceed the EWSR by 20\% and 10\%, respectively 
while the results for the $E2$ and $E3$ 
transitions agree well with the EWSR. 
For the $E4$ and $E5$ transitions, our results exhaust 
only 50\% of the EWSR. 
This may be due to the insufficient configuration space for
the $E4$ and $E5$ calculations.

On the other hand, there is no experimental constraint 
imposed on the nuclear electric current. 
However, the electromagnetic current should
satisfy the continuity equation required by charge 
conservation;
\begin{equation}
{\partial\over\partial t}\hat{\rho}_N+\nabla\hat{{\vec J}}_N=0.
\end{equation}
Using the nuclear Hamiltonian $H_N$, we rewrite it as
\begin{equation}
i[H_N,\hat{\rho}_N]+\nabla\hat{{\vec J}}_N=0.
\label{CCONS1}
\end{equation}
Taking the matrix element of Eq.~(\ref{CCONS1}) between 
the initial and the final nuclear states, 
we obtain charge conservation condition;
\begin{equation}
i\omega_N\rho_{N\lambda}^{I'I}(r)=
-\sqrt{\lambda\over{2\lambda+1}}
({d\over dr}-{{\lambda-1}\over r})
J_{N\lambda\lambda-1}^{I'I}(r)
+\sqrt{{\lambda+1}\over{2\lambda+1}}({d\over dr}
+{{\lambda+2}\over r})
J_{N\lambda\lambda+1}^{I'I}(r).
\label{CCONS2}
\end{equation}
Here, $\omega_N=E_{I'}-E_I$ is the energy of nuclear excitation.

For any kind of model calculation involving the nuclear current, 
it is necessary for the model to satisfy the charge 
conservation condition in order to observe the gauge
invariance\cite{FR74}. Unfortunately, the charge-current densities
constructed from the present RPA calculation do not 
satisfy the charge conservation of Eq.~(\ref{CCONS2}). 
The violation of the charge
conservation comes from the inconsistency of using empirical
single-particle energies together with the impulse charge-current
operators, as is discussed in Refs.~\cite{TA94,HO97}. 
It is desirable if one could construct a microscopic 
self-consistent model together with the nuclear current 
satisfying the charge conservation
which is realistic enough to reproduce the observed spectra and
$B(E\lambda)$ values.
However, the refinement of calculation will be left 
for a future work and at present we are satisfied 
with the fact that the calculated NP energies show only 
a small gauge violation even though
the empirical single-particle energies are used 
in the RPA calculation.

\section{Results and Discussion}

NP energy shifts are obtained by computing an 
energy shift for each of the RPA excitations and summing 
the results. Our calculation 
gives 38, 129,
160, 222, 202, 218 and 70 nuclear states for the $0^+$, $1^-$,
$2^+$, $3^-$, $4^+$, $5^-$ and $1^+$ excitations, respectively. 
Figure~3 shows the NP
energy spectra of the $1s_{1/2}$ state for the respective nuclear
spin-parities.
These spectra are calculated in the Coulomb gauge with the Coulomb
and transverse parts of the electromagnetic interaction. They
are very similar to the RPA spectra of $B(E\lambda)$ and $B(M1)$.

Table II summarizes the NP energies 
of the $1s_{1/2}$ state for
$^{208}_{~82}$Pb$^{81+}$. 
The 1st column denotes nuclear
spin-parities. 
The entries in the 2nd column indicate 
the contributions 
to the NP energy
from the ladder, cross and seagull terms 
as well as those of the
positive and negative energy intermediate-states of the electron.
The 3rd column shows the NP energies in the Feynman gauge, 
while the 4th column shows the NP energies in the Coulomb gauge. 
The transverse contributions are included in both columns. 
The 5th column shows the
Coulomb NP energies without the transverse contribution 
(hereafter referred to as CNP).  
The 6th column shows the results  of the  
previous NP calculation in the
Feynman gauge assuming a collective model for the nuclear
excitations\cite{YA01}. The 7th column is the CNP 
of the same calculation. Finally, the last column shows the CNP 
calculated by another group \cite{NE96}.

In Table II, it should be noted that the 
positive and negative energy continuum states 
of the electron always contribute to a NP energy with
opposite sign, the fact that is already 
observed in Refs.\cite{PL89,YA01}. 
It is also understood from the fact that 
the contribution of the negative energy electron 
describes the blocking of the response of
vacuum due to the occupied state of the atomic electron.  
In electronic atoms the virtual pair creation requires 
energy of only $2m_e$, which is smaller
than the typical excitation energy of a nucleus and is 
correspondingly important.

The 5th column  (the present CNP) may be compared with
the 7th and 8th columns of the previous calculations. All three
calculations show very good agreement with one another. 
The agreement indicates that provided the $B(E\lambda)$ 
values are similarly chosen, 
CNP of electronic atoms is not very sensitive 
to the detail of transition charge densities. 
The monopole NP energy shows some difference between the 
calculations. It is --4.0 meV (5th column), --7.2 meV 
(7th column) and --3.3 meV (8th column), respectively. 
The difference is mainly caused by the Dirac-electron wave 
functions used; those for the
finite charge distribution are used in the 5th and 8th columns, 
while those for a point charge are used in the 7th column. The
difference is conspicuous only for the monopole NP energy 
because the nuclear monopole transition-potential exists 
only inside the nucleus where the electron-wave functions 
generated from the point charge and the finite charge 
distribution differ appreciably.

The dipole NP energy also shows some difference, i.e., it is
--20.3 meV (the 4th column), --19.5 meV (the 6th column) 
and --17.6 meV (the 7th column), respectively. 
The difference between the 4th and the 7th
columns arises from the fact that the 
energy-weighted sum of $B(E1)$ over the RPA states 
exceeds the classical EWSR value by 10\% (see
Table I).

The NP energies including the transverse effect (the 3rd column) 
can be compared with the results of
Ref.~\cite{YA01} (the 6th column), both 
of which were calculated in the Feynman gauge. 
Agreement between the two calculations is  good
except for the nuclear monopole excitation. Here again,
we may conclude that NP
energies of heavy electronic atoms are not very sensitive to the
detail of transition current densities.

An important feature of the present calculation, 
which is in fact crucial 
for the numerical estimate of NP energies, is that there
exists a large violation of gauge invariance in the NP 
energy-shifts as far as only the
ladder and cross diagrams of Figs.~1(a) and 1(b) are taken into
account. By using the minimal prescription, 
the nonrelativistic electromagnetic
interaction involves the square of vector potential 
called a seagull term coming from the kinetic energy, 
and this term is necessary for
the gauge invariance in a nonrelativistic system \cite{FR74,RO83}. 
It is interesting to investigate whether 
the inclusion of the seagull term restores the gauge 
invariance of the present RPA calculation. 
(A proof of the gauge invariance is given in the Appendix.)

As is seen from Table II, this is nicely confirmed numerically. 
With the ladder and cross diagrams, the NP
energy for the $1s_{1/2}$ was +1.5 meV in the Feynman gauge, 
while --32.7 meV in the Coulomb gauge. 
The gauge dependence was 34.2 meV. However, after the 
inclusion  of the seagull term, it is --38.2 meV in
the Feynman gauge and --37.0 meV in the Coulomb gauge. The gauge
dependence is reduced to 1.2 meV.
This small gauge dependence shows that the seagull term 
is quite important in restoring the gauge 
invariance of the NP calculation. 
The fact also implies that the use of empirical 
single-particle energies in the RPA calculation does not 
introduce a serious violation of 
gauge invariance into the NP energies.

In Table II, we recognize that the nuclear dipole states give
predominant NP contributions. The transverse contribution in
particular is very large if the contribution from the 
positive and negative energy states are separately considered. 
For example, in the Coulomb gauge, the Coulomb plus 
transverse contribution from the ladder and cross 
diagrams is --120.5 meV, while the corresponding
contribution without the transverse term is --37.0 meV. 
The reason for this fact is understood as follows.  
For simplicity, we disregard the retardation of the 
interaction. Then Eq.~(\ref{DCoulomb}) gives
the Coulomb-Breit potentials for electron:
\begin{eqnarray}
V_{C}(\vec{r})_{I'I}&=&\int d\vec{r}'\frac{\rho_N(\vec{r}')_{I'I}}
{|\vec{r}-\vec{r}'|}, 
\label{CBP0}
\\
V_{T}(\vec{r})_{I'I}&=&\int d\vec{r}'
\frac{\vec{J}_N^T(\vec{r}')_{I'I}}{|\vec{r}-\vec{r}'|}.
\label{CBP}
\end{eqnarray}
In Fig.~4 the radial parts of the potentials due to the 1$^-$, 
14.6 MeV state are plotted after multipole expansions of
Eqs.~(\ref{CBP0}) and (\ref{CBP}). 
The Coulomb potential is larger than both of the 
transverse potentials in the region r $<$ 60 fm, 
while the transverse potential with $\lambda = 1, L = 0$
becomes larger than the Coulomb potential in the 
region r $>$ 60 fm. 
This long-range nature of the transverse potential explains the
fact that the $E1$ transverse contribution is dominant 
for electronic atoms because the Bohr radius of 
the $1s_{1/2}$ state is much larger than 60 fm. 
Since the radial dependence of the Coulomb potential 
(transverse potential) with  multipolarity $\lambda$ is 
$1/r^{\lambda+1}$ ($1/r^{\lambda}$) outside the nuclear region, 
the magnitude of the potentials in the region of the Bohr radius 
decreases as the multipolarity increases. 
Therefore, for multipoles other than the dipole, i.e., 
$\lambda^{\pi}$=$2^+$, $3^-$, $4^+$
and $5^-$, 
the NP energy is  mainly determined 
by the overlap between the electron transition density
and nuclear multipole potential in the region of the nuclear
surface, hence 
the transverse contribution is small
compared with the Coulomb contribution. 
The fact also explains that the Coulomb
contribution is predominant for heavy muonic atoms in any 
multipolarities because the overlap near the nucleus is 
always significant due to the small muon Bohr-radius
\cite{TA94}.

The dominance of the $E1$ contribution in
the NP energy can be seen more
clearly in the spectral density of the NP 
contribution from a particular nuclear 
excitation  as a function
of electron energy. 
In Fig.~5, 
these spectral functions in the Coulomb gauge are
shown for three different nuclear states; (a) $0^+$ (13.3 MeV), 
(b) $1^-$ (14.6 MeV) and (c) $2^+$ (10.2 MeV). 
In each panel, the solid
line shows the spectral function including 
the transverse contribution,
while the dotted line shows the result without the transverse
effect. 
The NP energy due to each of the nuclear states in Fig.~3 
is given by the integral of the corresponding spectral function over 
the electron energy.
One can see that the low energy region of the spectral 
function for the $1^-$ state (Fig.~5(b)) is 
different from the other two. The $E1$
spectrum shows a peak at threshold ($E_{i'}=m_e$). It should be
reminded that only the $<I'\parallel T_{10}(q)\parallel I>$ and
$<i'\parallel t_{10}(q)\parallel i>$ are nonvanishing at $q=0$, so
that the large overlap between these transverse form factors is
guaranteed in the low momentum region.

For the excited $L$-shell electrons, we can repeat 
the discussion that the transverse $E1$ multipole 
plays a crucial role in the NP effects of
hydrogenlike atoms and essentially determines the magnitude 
of the NP effects. 
The total NP shifts for the $1s_{1/2}$, $2s_{1/2}$,
$2p_{1/2}$, and $2p_{3/2}$ states are summarized in Table III 
and are compared with the Coulomb NP energies. 
The relative importance of the transverse contribution 
is about 10-20\% effect for the $1s_{1/2}$ and
$2s_{1/2}$ states, and the level shifts for the $2p_{1/2}$ and
$2p_{3/2}$ states are negligible.

Since the nuclear dipole states have predominant NP contributions, 
we must note here effects of the spurious 
center-of-mass motion of the nucleus on the NP energies. 
The present Migdal force brings down the lowest $1^-$ state 
to the imaginary eigenvalue of 1.32$i$ MeV. 
Since this $1^-$ state carries most of the 
spurious center-of-mass motion, we excluded this nuclear 
state from the NP calculation. 
The 0.7\% of the spurious center-of-mass motion remains 
in the rest of the 1$^-$ states, whose effects on the NP 
energies are negligible. 
Thus our results for NP energy 
due to the cross and ladder diagrams
contain intrinsic excitations only. 
On the other hand, the seagull contributions calculated by 
using Eq.~(\ref{ML})
contain both of the effects of intrinsic excitations 
and center-of-mass motion.
The seagull contribution coming from the center-of-mass 
motion must be eliminated
for the dipole mode.
This was achieved by using the effective dipole charges
$e_p=N/A$, $e_n=-Z/A$ instead of using the true charges 
$e_p=1$, $e_n=0$.

\section{summary}

We have calculated the NP energy shifts for the
hydrogenlike $^{208}_{~82}$Pb$^{81+}$ 
taking into account the effects of the electron
in the negative energy continuum besides the usual
contributions of the electron excited into higher unoccupied
orbitals. The evaluation of the NP energies 
contains the seagull graph as
well as the ladder and cross diagrams. The Dirac-electron wave 
functions were solved in the Coulomb potential with
a finite charge distribution for the nuclear ground state and 
the RPA wave functions were employed
for the nuclear excited states.

The results presented in the previous section can be summarized
as follows.
1)In the Feynman
(Coulomb) gauge, we obtained the NP energies of --38.2(--37.0),
--6.7(--6.4), --0.2(--0.2) and +0.0(+0.0) meV for the $1s_{1/2}$,
$2s_{1/2}$, $2p_{1/2}$ and $2p_{3/2}$ states, respectively. 
2) The nuclear dipole states have predominant contributions to 
the NP
energy of heavy electronic atoms. In particular the exchange of
transverse photon plays a crucial role. The net E1 contribution, 
however, becomes quite small due to the cancellation between
the contributions of the electron in the positive and 
negative energy intermediate states. 
3) Due to the cancellation, 
the transverse contribution to the total NP energy shift 
is about 10\% of the
Coulomb contribution for the $1s_{1/2}$ state. 
4) The NP shifts of electronic atoms have serious gauge 
dependence if one calculates them with only the ladder 
and cross diagrams of the
two-photon exchange processes. Indeed the NP energies 
for the $1s_{1/2}$
state due to these diagrams are +1.5 meV and --32.7 meV 
in the Feynman and Coulomb gauges, respectively. 
The inclusion of the seagull graph gives
the NP energies of --38.2 meV and -37.0 meV. The seagull
graph is  quite important in restoring the gauge invariance 
of the NP calculation. 
5) The NP energy for electronic atoms is not sensitive 
to the details of the transition charge
and current densities because 
it is primarily given by the overlap between 
an electron transition density and 
a nuclear multipole potential in the region outside the
nucleus. 
6) The present RPA calculation using the empirical
single-particle energies together with the impulse 
charge-current does not seem to introduce serious gauge 
dependence into the NP calculation.

\vskip30pt

\begin{acknowledgments}
We would like to thank N. Yamanaka and A. Ichimura for 
arousing our interest in the electronic-atom NP calculation.

\end{acknowledgments}

\vfill\eject

\appendix
\section{gauge invariance}

To prove the gauge invariance of the NP energy, 
we write the sum of the
ladder and cross contributions as follows,
\begin{eqnarray}
\Delta E_{NP}^L+\Delta E_{NP}^X&=&
-i \frac{(4\pi\alpha)^2}{2}\int \frac{d\omega}{2\pi}\int
\frac{d\vec{q}}{(2\pi)^3}
\int \frac{d\vec{q'}}{(2\pi)^3}
\nonumber
\\
&&
\Pi_e^{\mu\nu}(\omega,\vec{q},\vec{q}')
D_{\mu\xi}(\omega,q)D_{\zeta\nu}(\omega,q')
\Pi_N^{\xi\zeta}(\omega,\vec{q},\vec{q}'),
\label{NP}
\end{eqnarray}
where $\Pi_e^{\mu\nu}(\omega,\vec{q},\vec{q}')$ and
$\Pi_N^{\xi\zeta}(\omega,\vec{q},\vec{q}')$ are the electronic 
and nuclear polarization tensors defined by 
\begin{eqnarray}
\Pi_e^{\mu\nu}(\omega,\vec{q},\vec{q}')&=&
\sum_{i'}\left(
\frac{j_e^{\mu}(-\vec{q})_{ii'}j_e^{\nu}(\vec{q}')_{i'i}}
{\omega+\omega_e-i E_{i'}\epsilon}
-\frac{j_e^{\nu}(\vec{q}')_{ii'}j_e^{\mu}(-\vec{q})_{i'i}}
{\omega-\omega_e+i E_{i'}\epsilon}\right),
\label{lepton} \\
\Pi_N^{\xi\zeta}(\omega,\vec{q},\vec{q}')&=&
\sum_{I'}\left(
\frac{J_N^{\xi}(\vec{q})_{II'}J_N^{\zeta}(-\vec{q}')_{I'I}}
{\omega-\omega_N+i \epsilon}
-\frac{J_N^{\zeta}(-\vec{q}')_{II'}J_N^{\xi}(\vec{q})_{I'I}}
{\omega+\omega_N-i \epsilon}\right).
\label{hadron}
\end{eqnarray}
Photon propagators in the Feynman and Coulomb gauges are related
to each other by
\begin{eqnarray}
D^C_{\mu\xi}(q,\omega)=D^F_{\mu\xi}
(q,\omega)-\frac{1}{q^2+i \epsilon}
\left(\frac{q_{\mu}q_{\xi}-\omega(q_{\mu}g_{\xi 0}
+q_{\xi}g_{\mu 0})}{q^2}\right).
\label{DFC}
\end{eqnarray}
If both $q_{\mu}\Pi_e^{\mu\nu}= 0 $ and 
$q_{\xi }\Pi_N^{\xi\zeta}= 0 $ are satisfied, 
it is easy to see that the Feynman and Coulomb
gauges give the same result for the NP contributions  given by
Eq.~(\ref{NP}).

Multiplying both sides of the electron polarization tensor  
by $q_\mu$, and using the continuity equation of the
charge conservation,  one obtains 
\begin{eqnarray}
q_{\mu}\Pi_e^{\mu\nu}(\omega,\vec{q},\vec{q}')&=&
\sum_{i'}(<i|\hat{\rho}_e(-\vec{q})|i'><i'|\hat{j}^{\nu}_e
(\vec{q'})|i>
-<i|\hat{j}^{\nu}_e(\vec{q'})|i'><i'|\hat{\rho}_e(-\vec{q})|i>)
\nonumber
\\
&=&<i|[\hat{\rho}_e(-\vec{q}),\hat{j}^{\nu}_e(\vec{q'})]|i>.
\label{commutator}
\end{eqnarray}
In deriving the second equality, 
we have assumed the completeness of the intermediate states of the
electron. For the
electromagnetic charge and current operators
\begin{eqnarray}
\hat{\rho}_e(-\vec{q})=\left( 
\begin{array}{cc}
1&0\\
0&1
\end{array} \right) e^{i \vec{q}\cdot\vec{r}} ,
\hskip1cm
\hat{j}_e(\vec{q}')=\left( 
\begin{array}{cc}
0&\vec{\sigma}\\
\vec{\sigma}&0
\end{array} \right) e^{-i \vec{q}'\cdot\vec{r}},
\end{eqnarray}
used with the Dirac-electron wave functions, the commutation 
relation in Eq.~(\ref{commutator}) vanishes. 
Hence the gauge invariance  
$q_\mu \Pi_e^{\mu\nu}(\omega,\vec{q},\vec{q}') = 0$ for 
the electronic polarization tensor follows.

For the nuclear polarization tensor, 
we can obtain the similar form to Eq.~(\ref{commutator}) by
assuming the charge conservation as well as the 
completeness relation. In the present calculation, 
the impulse charge and current operators
\begin{eqnarray}
\hat{\rho}_N(-\vec{q})=\sum_i^{Z}e^{i\vec{q}\cdot\vec{r}_i} ,
\hskip1cm
\hat{J}_N(\vec{q})=\sum_i^{Z}\frac{
\overrightarrow{{\bf \nabla}}_{{\bf r}_i}
- \overleftarrow{{\bf \nabla}}_{{\bf r}_i}}{2m_p i }
e^{i \vec{q}\cdot\vec{r}_i}
+ \sum_i^{A}\left({\bf \nabla}_{\vec{r}_i}\times \vec{\mu}
\right)
e^{i \vec{q}\cdot\vec{r}_i},
\label{NOP}
\end{eqnarray}
are employed with the nonrelativistic RPA calculation. The
spin-current operator in the second term of $\hat{J}_N(q)$ 
commutes with
$\hat{\rho}_N$. Hence the spin current introduces no gauge
violation into the NP calculation of Eq.~(\ref{NP}). 
The convection-current,
on the other hand, does not commute with $\hat{\rho}_N$ leading to 
a violation of gauge invariance:
\begin{eqnarray}
<I|[\hat{\rho}_N(-\vec{q}),\hat{j}^{\nu}_N(\vec{q'})]|I>
=\frac{\vec{q}}{m_p}\rho_N(\vec{q}-\vec{q}')_{II}.
\label{eq:sggauge}
\end{eqnarray}
Therefore the NP contribution given by Eq.~(\ref{NP}) is not gauge
invariant with the impulse charge-current operators. When the seagull
tensor of  Eq.~(\ref{SEA}) is added to 
the nuclear polarization tensor of Eq.~(\ref{hadron}), 
this term is just canceled. Hence the gauge invariance of
NP calculation is restored by the seagull term
together with the the ladder and cross terms.

\clearpage
\setlength{\baselineskip}{0.0in}

TABLE I. The energy-weighted sums of $B(E\lambda)$ over 
the RPA states. The classical EWSR values \cite{BM75} 
are also shown for comparison. 
The values are given in units of e$^2$b$^{\lambda}\cdot$MeV\\
\vspace{1mm}\\
\begin{tabular}{ccc}
\hline\hline
E$\lambda$&
\hspace{4mm}Present calculation\hspace{4mm}&
\hspace{4mm}Classical EWSR $^{a}$\hspace{4mm}\\
\hline
$E0$$^{b}$ & 1.97 & 1.64 \\
$E1$$^{c}$ & 8.15 & 7.38 \\
$E2$       & 22.2 & 20.5 \\
$E3$       & 24.4 & 23.4 \\
$E4$       & 14.2 & 23.6 \\
$E5$       & 11.3 & 23.1 \\
$M1$$^{d}$ & 294 \\
\hline\hline
\end{tabular}

\begin{description}
\item{$^{a}$}
The radial moments $<r^{\lambda}>_p$ in the classical EWSR are 
calculated by the Fermi charge distribution (\ref{FermiD}).
\item{$^{b}$}
The $E0$ operator is defined as $O(E0)$=$\sum_p r^2/\sqrt{4\pi}$.
\item{$^{c}$}
The $E1$ operator is defined as $O(E1)$=$\sum_i -1/2\tau_3rY_{1\mu}$.
\item{$^{d}$}
The value given in units of $\mu_{\rm N}\cdot$ MeV.
\end{description}

\clearpage
TABLE II. Nuclear polarization correction (meV) of the 
$1s_{1/2}$ state in $^{208}_{~82}$Pb$^{81+}$. Energy shifts
$\Delta E^L$, $\Delta E^X$ and $\Delta E^{SG}$ 
are contributions of the ladder, cross and seagull terms,
respectively, while $\Delta E^+$ ($\Delta E^-$) denotes 
contribution from the positive (negative) energy 
intermediate states of the electron.
\\
\vspace{1mm}\\
\begin{math}
\begin{array}{llrrrrrr}
\hline\hline
&&
{\rm present}^{a}&
{\rm present}^{b}&
{\rm present}^{c}&
\hspace{3mm}{\rm Ref.~\cite{YA01}}^{d}&
\hspace{3mm}{\rm Ref.~\cite{YA01}}^{e}&
\hspace{3mm}{\rm Ref.~\cite{NE96}}^{f}\\
\lambda^{\pi}\hspace{5mm}&{\rm Contribution}&
\hspace{3mm}{\rm Feynman(NP)}&
\hspace{3mm}{\rm Coulomb(NP)}&
\hspace{8mm}{\rm CNP}&
\hspace{8mm}{\rm NP}&
\hspace{8mm}{\rm CNP}&
\hspace{8mm}{\rm CNP}\\
\hline\\
0^+
& {\Delta E^{L+}}
& -5.7
& -6.5
& -7.0\\
& {\Delta E^{L-}}
& +0.4
& +0.2\\
& {\Delta E^{X+}}
& -1.2
& -0.2\\
& {\Delta E^{X-}}
& +2.7
& +2.7
& +3.0\\
& {\Delta E^L\!+\!\Delta E^X }
& -3.8
& -3.9
& -4.0
& -6.6
& -7.2
& -3.3\\
& {\Delta E^{SG+}}
& +0.7
&  0.0\\
& {\Delta E^{SG-}}
& -0.9
&  0.0\\
& {\Delta E^L\!+\!\Delta E^X\!+\!\Delta E^{SG} }
& -3.9
& -3.9\\
\vspace{0mm}\\
1^-
& {\Delta E^{L+}}
& -119.1
& -91.1
& -37.0\\
& {\Delta E^{L-}}
& +74.0
& +37.6\\
& {\Delta E^{X+}}
& -49.8
& -29.4\\
& {\Delta E^{X-}}
& +110.1
& +64.2
& +16.7\\
& {\Delta E^L\!+\!\Delta E^X }
& +15.2
& -18.7
& -20.3
& +16.3
& -19.5
& -17.6\\
& {\Delta E^{SG+}}
& +144.2
& +87.8\\
& {\Delta E^{SG-}}
& -186.5
& -95.1\\
& {\Delta E^L\!+\!\Delta E^X\!+\!\Delta E^{SG}}
& -27.1
& -26.0\\
\vspace{0mm}\\
2^+
& {\Delta E^{L+}}
& -13.0
& -15.3
& -14.4\\
& {\Delta E^{L-}}
& +2.0
& +0.4\\
& {\Delta E^{X+}}
& -3.2
& -0.5\\
& {\Delta E^{X-}}
& +8.1
& +9.1
& +8.6\\
& {\Delta E^L\!+\!\Delta E^X }
& -6.1
& -6.3
& -5.8
& -7.0
& -6.3
& -5.8\\
& {\Delta E^{SG+}}
& +4.4
& +2.7\\
& {\Delta E^{SG-}}
& -4.1
& -2.1\\
& {\Delta E^L\!+\!\Delta E^X\!+\!\Delta E^{SG}}
& -5.7
& -5.7\\
\vspace{0mm}\\
3^-
& {\Delta E^{L+}}
& -5.0
& -6.5
& -6.3\\
& {\Delta E^{L-}}
& +1.0
& +0.1\\
& {\Delta E^{X+}}
& -1.6
& -0.1\\
& {\Delta E^{X-}}
& +3.2
& +4.1
& +4.0\\
& {\Delta E^L\!+\!\Delta E^X }
& -2.4
& -2.4
& -2.3
& -2.9
& -2.6
& -2.6\\
& {\Delta E^{SG+}}
& +1.0
& +0.6\\
& {\Delta E^{SG-}}
& -0.8
& -0.4\\
& {\Delta E^L\!+\!\Delta E^X\!+\!\Delta E^{SG}}
& -2.2
& -2.2\\
\vspace{0mm}\\
4^+
& {\Delta E^{L+}}
& -1.1
& -1.5
& -1.4\\
& {\Delta E^{L-}}
& +0.2
&  0.0\\
& {\Delta E^{X+}}
& -0.4
&  0.0\\
& {\Delta E^{X-}}
& +0.7
& +0.8
& +0.8\\
& {\Delta E^L\!+\!\Delta E^X}
& -0.7
& -0.7
& -0.6\\
& {\Delta E^{SG+}}
& +0.5
& +0.4\\
& {\Delta E^{SG-}}
& -0.4
& -0.3\\
& {\Delta E^L\!+\!\Delta E^X\!+\!\Delta E^{SG}}
& -0.6
& -0.6\\
\vspace{0mm}\\
\hline
\end{array}
\end{math}

\clearpage

TABLE II. $(Continued)$.
\\
\vspace{1mm}\\
\begin{math}
\begin{array}{llrrrrrr}
\hline\hline
&&
{\rm present}^{a}&
{\rm present}^{b}&
{\rm present}^{c}&
\hspace{3mm}{\rm Ref.~\cite{YA01}}^{d}&
\hspace{3mm}{\rm Ref.~\cite{YA01}}^{e}&
\hspace{3mm}{\rm Ref.~\cite{NE96}}^{f}\\
\lambda^{\pi}\hspace{5mm}&{\rm Contribution}&
\hspace{3mm}{\rm Feynman(NP)}&
\hspace{3mm}{\rm Coulomb(NP)}&
\hspace{8mm}{\rm CNP}&
\hspace{8mm}{\rm NP}&
\hspace{8mm}{\rm CNP}&
\hspace{8mm}{\rm CNP}\\
\hline\\
5^-
& {\Delta E^{L+}}
& -0.4
& -0.5
& -0.4\\
& {\Delta E^{L-}}
& +0.1
&  0.0\\
& {\Delta E^{X+}}
& -0.1
&  0.0\\
& {\Delta E^{X-}}
& +0.2
& +0.2
& +0.2\\
& {\Delta E^L\!+\!\Delta E^X}
& -0.2
& -0.2
& -0.2\\
& {\Delta E^{SG+}}
& +0.3
& +0.1\\
& {\Delta E^{SG-}}
& -0.2
& -0.1\\
& {\Delta E^L\!+\!\Delta E^X\!+\!\Delta E^{SG}}
& -0.2
& -0.2\\
\vspace{0mm}\\
1^+
& {\Delta E^{L+}}
& -0.4
& -0.4\\
& {\Delta E^{L-}}
& +0.1
& +0.1\\
& {\Delta E^{X+}}
& -0.2
& -0.2\\
& {\Delta E^{X-}}
& +0.1
& +0.1\\
& {\Delta E^L\!+\!\Delta E^X}
& -0.4
& -0.4\\
& {\Delta E^{SG+}}
& +3.8
& +3.8\\
& {\Delta E^{SG-}}
& -1.7
& -1.7\\
& {\Delta E^L\!+\!\Delta E^X\!+\!\Delta E^{SG}}
& +1.7
& +1.7\\
\vspace{1mm}\\
\hline
\vspace{0mm}\\
{\rm Total}\\
& {\Delta E^{L+}}
& -144.7
& -121.8
& -66.5\\
& {\Delta E^{L-}}
& +77.8
& +38.3\\
& {\Delta E^{X+}}
& -56.7
& -30.4\\
& {\Delta E^{X-}}
& +125.1
& +81.2
& +33.3\\
& {\Delta E^L\!+\!\Delta E^X}
& +1.5
& -32.7
& -33.2
& -0.2
& -33.6
& -29.3\\
& {\Delta E^{SG+}}
& +154.9
& +95.4\\
& {\Delta E^{SG-}}
& -194.6
& -99.7\\
& {\Delta E^L\!+\!\Delta E^X\!+\!\Delta E^{SG}}
& -38.2
& -37.0\\
\vspace{1mm}\\
\hline\hline\\
\end{array}
\end{math}

\begin{description}
\item{$^{a}$}
The NP energies in the Feynman gauge.
\item{$^{b}$}
The NP energies in the Coulomb gauge.
\item{$^{c}$}
The unretarded NP energies in the Coulomb gauge.
\item{$^{d}$}
The NP energies evaluated in the Feynman gauge. 
Electron wave functions were solved by assuming 
the point charge for the nuclear ground state. 
Nuclear transition-charge densities were determined by 
a collective model. 
They were normalized to the observed $B(E\lambda)$
for the low-lying nuclear states and EWSR values for the 
high-lying giant resonances. 
Nuclear current densities $J_{N\lambda\lambda-1}(r)$
were obtained by solving the equation of the charge conservation 
(\ref{CCONS2}) assuming $J_{N\lambda\lambda+1}(r)=0$.
\item{$^{e}$}
Same as  $d$ except for the unretarded NP energies.
\item{$^{f}$}
The unretarded NP energies. Same as $e$ except for 
the electron wave functions solved by assuming a 
finite charge distribution for the nuclear ground state. 
\end{description}

\clearpage

TABLE III. Total nuclear polarization (meV) of the $1s_{1/2}$,
$2s_{1/2}$, $2p_{1/2}$, and $2p_{3/2}$ states 
of $^{208}_{~82}$Pb$^{81+}$ both in the Feynman and 
Coulomb gauges. 
The abbreviation CNP denotes the unretarded results.
\\
\vspace{1mm}\\
\begin{math}
\begin{array}{lccc}
\hline\hline
{\rm States}&
\hspace{5mm}{\rm Feynman(NP)}\hspace{5mm}&
\hspace{5mm}{\rm Coulomb(NP)}\hspace{5mm}&
\hspace{5mm}{\rm CNP}\hspace{5mm}\\
\hline
{\rm 1s}_{1/2}&
-38.2&
-37.0&
-33.2\\
{\rm 2s}_{1/2}&
-6.7&
-6.4&
-5.7\\
{\rm 2p}_{1/2}&
-0.2&
-0.2&
-0.6\\
{\rm 2p}_{3/2}&
+0.0&
+0.0&
-0.0\\
\hline\hline
\end{array}
\end{math}

\clearpage
\setlength{\baselineskip}{0.3in}
{\large{\bf Figure captions}}
\\

FIG.~1.
Diagrams contributing to nuclear polarization in lowest order; 
(a) ladder diagram, (b) cross diagram, (c) seagull diagram.
\\

FIG.~2.
Electronic Coulomb form-factors 
$<E_{i'},p_{1/2}\parallel m_1(q)\parallel 1s_{1/2}>$
for the $E_{i'}$=2, 6 and 10 MeV states (solid line). 
The dotted line is the nuclear Coulomb form factor 
for the 14.6 MeV 1$^-$  state.
\\

FIG.~3.
Nuclear polarization spectra as functions of nuclear excitation
energy. The Coulomb gauge was assumed.
\\

FIG.~4.
The transition potentials due to the nuclear 1$^-$, 14.6 MeV state.
The Coulomb-Breit propagator was assumed. The solid, dotted and
dash-dotted lines denote the potentials corresponding to $\rho_1$(r),
$J_{10}^T$(r) and $J_{12}^T$(r), respectively. The transverse
potential  $J_{10}^T$(r) is seen to be larger than the Coulomb
potential in the region $r>60$ fm.
\\

FIG.~5.
Spectral densities of nuclear polarization for (a) 0$^+$, 
13.3 MeV, (b) 1$^-$, 14.6 MeV and (c) 2$^+$, 10.2 MeV states 
as functions of electron energy $E_{i'}$. 
The solid line denotes calculation with both
the Coulomb and transverse parts of the electromagnetic 
interaction, while the dotted line denotes calculation 
with only the Coulomb part of the
interaction. Electron intermediate states, (a) $|E_{i'},s_{1/2}>$,
(b) $|E_{i'},p_{1/2}>$ and (c) $|E_{i'},d_{3/2}>$
are assumed in the respective panels.

\end{document}